\pdfoutput=1
\documentclass[prd,preprint,floats,floatfix,aps,nofootinbib,preprintnumbers]{revtex4-1}
\usepackage{graphics,graphicx,psfrag}
\usepackage{amsmath,amssymb}
\usepackage[sort&compress]{natbib}
\usepackage{bm}	
\usepackage{verbatim}
\usepackage{multirow}
\usepackage{subfigure}
\usepackage{ulem}
\usepackage{color}
\usepackage[linktocpage=true,bookmarksnumbered=true,colorlinks=true,plainpages,linkcolor=blue,citecolor=red]{hyperref}
\usepackage{soul}	
\soulregister\cite7
\soulregister\ref7
\soulregister\pageref7
\def\beq{\begin{equation}}
\def\eeq{\end{equation}}
\def\bea{\begin{eqnarray}}
\def\eea{\end{eqnarray}}
\def\bealn{\begin{eqnarray}}
\def\eealn{\end{eqnarray}}

\def\ifb{\rm fb^{-1}}

\def\tev{\rm TeV}
\def\zp{{Z^\prime}}	
\def\dcol{D_{\rm{col}}}

\begin{document}
\preprint{
{\vbox {
\hbox{\bf MSUHEP-140608}
\hbox{\today}
}}}
\vspace*{2cm}

\title{The Color Discriminant Variable and Scalar Diquarks at the LHC}
\vspace*{0.25in}   
\author{R. Sekhar Chivukula}
\email{sekhar@msu.edu}
\author{Pawin Ittisamai}
\email{ittisama@msu.edu}
\author{Kirtimaan Mohan}
\email{kamohan@pa.msu.edu}
\author{Elizabeth H. Simmons}
\email{esimmons@msu.edu}
\affiliation{\vspace*{0.1in}
Department of Physics and Astronomy\\
Michigan State University, East Lansing U.S.A.\\}
\vspace*{0.25 in} 


\begin{abstract}
\vspace{0.5cm}
\noindent
The LHC is actively searching for narrow dijet resonances corresponding to physics beyond the Standard Model.  Among the many resonances that have been postulated (e.g., colored vectors, scalars, and fermions) one that would have a particularly large production rate at the LHC would be a scalar diquark produced in the $s$-channel via fusion of two valence quarks. In previous work, we introduced a color discriminant variable that distinguishes among various dijet resonances, drawing on measurements of the dijet resonance mass, total decay width and production cross-section.  Here, we show that this model-independent method applies well to color-triplet and color-sextet scalar diquarks, distinguishing them clearly from  other candidate resonances.  We also introduce a more transparent theoretical formulation of the color discriminant variable that highlights its relationship to the branching ratios of the resonance into incoming and outgoing partons and to the properties of those partons.  While the original description of the color discriminant variable remains convenient for phenomenological use upon discovery of a new resonance, the new formulation makes it easier to predict the value of the variable for a given class of resonance.

\end{abstract}

\maketitle

\section{Introduction}
\label{sec:intro}
There have been numerous searches for Beyond-the-Standard-Model (BSM) resonances decaying to dijet final states at colliders including the CERN $\mathrm{S\bar{p}pS}$~\cite{Arnison:1986vk, Alitti:1993pn}, the Tevatron~\cite{Abe:1989gz, Abe:1995jz, Abe:1997hm, Abazov:2003tj, Aaltonen:2008dn}, and the Large Hadron Collider (LHC)~\cite{Aad:2010ae, Khachatryan:2010jd, Chatrchyan:2011ns, Aad:2011fq, atlas:2012nma,ATLAS:2012pu,ATLAS:2012qjz,Chatrchyan:2013qha, Khachatryan:2015sja}. As no new dijet resonances have been discovered so far, the current exclusion limits on the production cross section for those of sufficiently narrow width have been set by searches carried out by ATLAS and CMS collaborations at the LHC with a center-of-mass energy of $8 \,\tev$~\cite{ATLAS:2012pu, Chatrchyan:2013qha, CMS:kxa}. The upgraded, higher-energy LHC will be able to seek a resonance with a larger mass, and the greater integrated luminosity will enable the experiments to reach new discovery thresholds; see, for example,~\cite{Han:2010rf,Yu:2013wta} for recent studies of dijets at the future LHC.

When the LHC discovers a new dijet resonance, it will be crucial to determine the spin, color, and other properties of the resonance in order to understand what kind of BSM context it represents. In previous work~\cite{Atre:2013mja}, we introduced a way to distinguish whether a vector resonance is either a leptophobic color-singlet or a color-octet, using a construct that we called a ``color discriminant variable'', $\dcol$. The variable is constructed from the dijet cross-section for the resonance ($\sigma_{jj}$), its mass ($M$), and its total decay width ($\Gamma$), observables that will be available from the dijet channel measurements of the resonance\footnote{Note that $D_{col}$ is dimensionless in the units where $\hbar = c = 1$.}:
	\beq
		\dcol \equiv \frac{M^3}{\Gamma} \sigma_{jj},
	\label{eq:dcol}
	\eeq
For a narrow-width resonance, the color discriminant variable is independent of the resonance's overall coupling strength.

We have applied the color discriminant variable technique both to flavor universal vector resonances with  identical couplings to all quarks~\cite{Atre:2013mja} and also to more generic flavor non-universal vector resonances~\cite{Chivukula:2014npa} whose couplings to quarks vary by electric charge, chirality, or generation.  In the latter case, combining the color discriminant variable with  information from resonance decays to heavy top ($t\bar{t}$) or bottom ($b\bar{b}$) flavors still enables one to determine what type of resonance has been discovered.  We have also shown~\cite{Chivukula:2014pma} that the method can be used to separate fermionic or scalar dijet resonances from vector states.

In this work, we further extend the color discriminant variable technique in two directions. First, we re-frame the theoretical discussion of the variable in more general language that shows its broader applicability and its relationship to the properties of the partons involved in production and decay of a narrow resonance.  In addition, we show that $D_{col}$ can be used to distinguish a color-triplet or color-sextet scalar diquark (a weak-singlet state coupling to two quarks) from weak-singlet vector dijet resonances that couple to a quark/anti-quark pair, such as a coloron (color-octet) or $Z'$ (color-singlet).

The rest of the paper is organized as follows:  In the next section, we introduce the scalar diquark and vector resonances studied in this work.  Section~\ref{sec:coldis} reviews the original presentation of the color discriminant variable, provides a new theoretical framework for it, and applies it to diquarks.  Section~\ref{sec:paramspace} establishes the range of coupling and mass parameter space where the color discriminant analysis can be used for diquarks.  We show how $D_{col}$ separates diquarks from one another and from vector boson resonances in Section~\ref{sec:result-dcols} focusing on resonances with masses of $3-7$ TeV at the $\sqrt{s}=14$ TeV LHC with integrated luminosities up to $1000\, \ifb$.  The final section summarizes our conclusions.

\section{New resonances coupling to quarks}
\label{sec:genpar}
In this section, we introduce our parametrization of the couplings of scalar diquark resonances coupling to $qq$ and the vector resonances coupling to $q\bar{q}$.

\subsection{Scalar Diquarks}
\label{sec:diquark-intro}

The diquarks we consider are weak-singlet scalar resonances coupling to two {\it quarks}. Color triplet objects of this kind appear in the 27-dimensional representation of $E_6$ string-inspired grand-unified theories  \cite{Angelopoulos:1986uq,Hewett:1988xc,King:2005jy,Kang:2007ib}, and their hadron-collider phenomenology has been studied, for example, in Refs.~\cite{Katsilieris:1992fa,Cakir:2005iw,Bauer:2009cc,Gogoladze:2010xd}. Color triplet scalars are the canonical benchmark diquarks considered by CMS \cite{Harris:2011bh}.  The hadron collider phenomenology of color sextet scalars has been studied here \cite{Mohapatra:2007af,Chen:2008hh,Berger:2010fy}. In the absence of flavor symmetries \cite{Chivukula:1987py,Arnold:2009ay}, there are very strong constraints on the couplings of these particles~\cite{Ma:1998pi}. NLO corrections to diquark production \cite{Han:2009ya,Das:2015lna} and the showering of sextet particles \cite{Richardson:2011df} have been studied as well.

Before discussing specific diquarks in detail, it is important to specify how we will handle flavor issues.  There are several options.  One can simply assume that the couplings are written in the mass-diagonal basis, in terms of the appropriate left- and right-handed fields. While this is unnatural, it is
the stated (or unstated) assumption of most phenomenological work aside from papers specifically addressing flavor limits \cite{Ma:1998pi}.  Alternatively (as we have done), one can assign appropriate flavor quantum-numbers \cite{Chivukula:1987py} to the specific color-triplet or color-sextet states as discussed in \cite{Arnold:2009ay}. In this case, one has a full mass-degenerate flavor multiplet of any of the particles present (e.g., a flavor triplet of color-triplet diquarks coupling to right-handed up-type quarks as in Case V of \cite{Arnold:2009ay}). Remarkably, if one is studying {\it single}
production of a diquark from $qq$ annihilation at the LHC, the signal is essentially the same as obtained using the first strategy; since typically one partonic production mechanism dominates and each diquark carries the flavor(s) of the incoming quarks, only one such object is largely responsible for any given single production signal.

Note that, for color-triplet diquarks, in some cases {\it flavor anti-symmetry} is sufficient
to eliminate tree-level $\Delta F=2$ processes, and hierarchical Yukawa couplings patterned
on the SM yukawa matrices may be sufficient to eliminate dangerous contributions to flavor-changing
neutral-current processes \cite{Giudice:2011ak}.

\subsubsection{Color Triplet Diquarks}
\label{sec:triplets}

We will use the classification system from  \cite{Arnold:2009ay} to describe the different types of weak-singlet, color-triplet diquarks associated with a weak doublet quark field (${\bf\rm Q_L}$) and weak singlet quark fields (${\bf\rm u_R}$ and ${\bf\rm d_R}$) of any generation.  Keeping in mind that the color-triplet state of two quarks will be anti-symmetric in color and in Lorentz spinor indices, it will also need to be anti-symmetric under the combination of flavor and SU(2) indices.

A weak-singlet, charge $1/3$ diquark coupling to ${\bf\rm Q_L} {\bf\rm Q_L}$ will be anti-symmetric in $SU(2)$ indices, so it must be symmetric in flavor, which places it in a sextet representation of $SU(3)_{\bf\rm Q_L}$.  This corresponds to case XI in \cite{Arnold:2009ay}.  The six distinct states will fall into the flavor combinations: $(u_L d_L),\, (u_L s_L + c_L d_L)/\sqrt{2},\, (c_L s_L),\, (c_L b_L + t_L s_L)/\sqrt{2},\, (t_L b_L),\, (u_L b_L + t_L d_L)/\sqrt{2}$.  Only the first two are phenomenologically relevant for our study of dijet resonances at LHC; we will focus on the $u_L d_L \equiv \omega_3$ diquark state, which would be the most readily produced at LHC.

A charge $1/3$ diquark coupling to ${\bf\rm u_R}{\bf\rm d_R}$ will be anti-symmetric in flavor (since the right-handed quarks are weak singlets), corresponding to the $(3,3)$ representation of $SU(3)_{\bf\rm u_R} \times SU(3)_{\bf\rm d_R}$, as in case IX of  \cite{Arnold:2009ay}.  The nine flavor combinations will be $u_R d_R, u_R s_R, u_R b_R,$ and so on.  While the four states coupling only to first or second generation quarks could potentially be relevant for our study, we will focus on the $u_R d_R \equiv \tilde{\omega}_3$ state, which would be most readily produced at LHC.\footnote{The particles $\omega_3$ and $\tilde{\omega}_3$ here correspond to $D$ and $D^c$ in \cite{Angelopoulos:1986uq}.}

Similarly, a charge $4/3$ diquark coupling to ${\bf\rm u_R}{\bf\rm u_R}$ will also be anti-symmetric in flavor, corresponding to the $\bar{3}$ representation of $SU(3)_{\bf\rm u_R}$, as in case V of  \cite{Arnold:2009ay}.  Of the three flavor states making up the triplet $[u_R c_R,\, u_R t_R,\, c_R t_R]$ only the first is potentially relevant to our study, but the small charm parton distribution function will suppress its production rate.  The charge $-2/3$ diquark coupling to ${\bf\rm d_R} {\bf\rm d_R}$ (case VII of  \cite{Arnold:2009ay}) is analogous, with down-type flavors substituted everywhere for up-type flavors.  We will not discuss either of these further.

 Following the notation in~\cite{Han:2009ya}, we write the interactions of these diquark states with quarks as
 \begin{equation}
 {\cal L} = 2\sqrt{2}\left(\bar{K}_3\right)^{ab}_c\left[
 \lambda_\omega \omega_3^c \bar{u}_{La} d^C_{Rb} + \lambda_{\tilde{\omega}} \tilde{\omega}_3^c \bar{u}_{Ra} d^C_{Lb}\right]+h.c.~,
 \end{equation}
where $a,b$ and $c$ are color (triplet) indices, $\bar{K}_3$ is the color Clebsch-Gordan coefficient connecting~\cite{Han:2009ya} two triplets to an anti-triplet (related to $\epsilon_{abc}$), and $\lambda_{\omega,\tilde{\omega}}$ are unknown coupling constants. We then find the decay widths
\begin{equation}
\Gamma(\omega_3, \tilde{\omega}_3 \to u + d) = \frac{\lambda^2_{\omega,\tilde{\omega}}}{2\pi}
\,M_{\omega_3,\tilde{\omega}_3}~,
\label{eq:D-triplet-width}
\end{equation}
and, in the narrow-width approximation, the hadronic cross sections
\begin{equation}
\sigma(pp \to \omega_3, \tilde{\omega}_3) = \frac{2\pi}{3} \lambda^2_{\omega,\tilde{\omega}}
\left[\frac{1}{s} \frac{d{\cal L}^{ud}}{d\tau}\right]_{\tau=\frac{M^2_{\omega_3,\tilde{\omega}_3}}{s}}~,
\label{eq:D-triplet-xsec}
\end{equation}
where $M_{\omega,\tilde{\omega}}$ are the masses of the diquark resonances, $s$ is the $pp$ hadronic center of mass energy, and the luminosity function $d{\cal L}/d\tau$ is defined in Eq. (\ref{eq:lumi-fun}) .

\subsubsection{Color Sextet Diquarks}
\label{sec:sextets}

For color sextet scalar diquarks, we can also use the classification system in \cite{Arnold:2009ay} to enumerate the possible states.  Because the color-sextet state is symmetric in color and anti-symmetric in Lorentz spinor indices, it must be {\it symmetric} under the combination of flavor and SU(2) indices.

The weak-singlet, charge $1/3$ diquark coupling to ${\bf\rm Q_L} {\bf\rm Q_L}$ will be anti-symmetric in $SU(2)$ indices, so it must be anti-symmetric in flavor, placing it in a triplet representation of $SU(3)_{\bf\rm Q_L}$.  This corresponds to case XII in \cite{Arnold:2009ay}.  The three states will fall into the flavor combinations: $(u_L s_L - c_L d_L),\, (u_L b_L - t_L d_L),\, (c_L b_L - t_L s_L)$.  Only the first of these is potentially relevant for our work; we will follow \cite{Chen:2008hh} in denoting it as $\delta_6$.

There is also a charge 1/3 diquark coupling to ${\bf\rm u_R}{\bf\rm d_R}$ transforming under the symmetric $(3,3)$ representation of $SU(3)_{\bf\rm u_R} \times SU(3)_{\bf\rm d_R}$, as in case X of  \cite{Arnold:2009ay}.  The nine flavor combinations will be $u_R d_R,\, u_R s_R,\, u_R b_R,$ and so on.  While the four states coupling to first or second generation quarks could potentially be relevant for our study, we will focus on the $u_R d_R \equiv \Delta_6$ state \cite{Chen:2008hh}, which would be most readily produced at LHC.

The color-sextet, charge 4/3 diquark coupling to ${\bf\rm u_R}{\bf\rm u_R}$ will also be symmetric in flavor, corresponding to the $6$ representation of $SU(3)_{\bf\rm u_R}$, as in case VI of  \cite{Arnold:2009ay}.  Of the six flavor states [$(u_R u_R),\, (u_R c_R + c_R u_R)/\sqrt{2},$ etc.] the first is most relevant to our study; we will denote it $\Phi_6$, following \cite{Chen:2008hh}.   The corresponding charge -2/3 state composed of down-flavored quarks (case VIII of \cite{Arnold:2009ay}) will likewise be called $\phi_6 \equiv d_R d_R$, as in \cite{Chen:2008hh}.

Following the notation of Ref. ~\cite{Han:2009ya}, we may write the corresponding interactions between the diquarks and light-generation fermions as
\begin{equation}
{\cal L} = 2 \sqrt{2} (\bar{K}_6)^{ab}_\gamma \left[
\lambda_\delta \delta^\gamma_6 (\bar{u}_{La} s^C_{Rb}-\bar{c}_{La}d^C_{Rb})
+\lambda_\Phi \Phi^\gamma_6 \bar{u}_{Ra}u_{Lb} + \lambda_\phi \phi^\gamma_6 \bar{d}_{Ra}d_{Lb} + \lambda_\Delta \Delta^\gamma_6 \bar{u}_{Ra} d_{Lb}
\right] + h.c.~,
\end{equation}
where $a,b$ are triplet color indices and $\gamma$ is a sextet color index,
$\bar{K}_6$ is the Clebsch-Gordan coefficient connecting two $SU(3)$ triplets to a sextet, and $\lambda_{\delta, \Phi,\phi, \Delta}$ are unknown coupling constants. We then find the decay widths of the charge $1/3$ states to be
\begin{align}
\Gamma(\delta_6 \to u + s) & = \Gamma(\delta_6 \to c + d) = \frac{\lambda^2_\delta}{2\pi} M_{\delta_6}~,\\
\Gamma(\Delta_6 \to u + d) & = \frac{\lambda^2_\Delta}{2\pi} M_{\Delta_6}~,
\end{align}
while, due to identical final state particles, widths of the charge $4/3$ states are
\begin{align}
\Gamma(\Phi_6 \to u + u) & = \frac{\lambda^2_\Phi}{4\pi} M_{\Phi_6}~,\\
\Gamma(\phi_6 \to d + d) & = \frac{\lambda^2_\phi}{4\pi} M_{\phi_6}~.
\end{align}
The corresponding cross sections are
\begin{align}
\sigma(pp \to \delta_6) & = \frac{4\pi \lambda^2_\delta}{3}
\left[\frac{1}{s} \frac{d{\cal L}^{us}}{d\tau} + \frac{1}{s} \frac{d{\cal L}^{cd}}{d\tau}\right]_{\tau=\frac{M^2_{\delta_6}}{s}}\\
\sigma(pp \to \Delta_6) & = \frac{4\pi \lambda^2_\Delta}{3}
\left[\frac{1}{s} \frac{d{\cal L}^{ud}}{d\tau}\right]_{\tau=\frac{M^2_{\Delta_6}}{s}}\\
\sigma(pp \to \Phi_6) & = \frac{4\pi \lambda^2_\Phi}{3}
\left[\frac{1}{s} \frac{d{\cal L}^{uu}}{d\tau}\right]_{\tau=\frac{M^2_{\Phi_6}}{s}}\\
\sigma(pp \to \phi_6) & = \frac{4\pi \lambda^2_\phi}{3}
\left[\frac{1}{s} \frac{d{\cal L}^{dd}}{d\tau}\right]_{\tau=\frac{M^2_{\phi_6}}{s}}~.
\end{align}

\subsection{Vector Bosons}
\label{sec:vector-intro}

A color-octet vector boson (coloron) arises from extending the gauge group of the strong sector. The couplings between quarks and the color-octet can be either flavor universal or flavor non-universal, and either chiral or vectorial. Examples of flavor universal scenarios include the classic axigluon~\cite{Frampton:1987dn, Frampton:1987ut} and coloron~\cite{Chivukula:1996yr, Simmons:1996fz} where all quarks are assigned to the same $SU(3)$ group in the extended color gauge sector. Flavor non-universal scenarios appear in the case of the topgluon where the third generation quarks are assigned to one $SU(3)$ group and the light quarks to the other~\cite{Hill:1991at, Hill:1994hp}, and the newer axigluon models where different chiralities of the same quark can be charged under different groups~\cite{Antunano:2007da, Ferrario:2009bz, Frampton:2009rk, Rodrigo:2010gm, Chivukula:2010fk, Tavares:2011zg}. Other examples of color-octet vector bosons include excited gluons in extra-dimensional models (Kaluza-Klein gluons \cite{Dicus:2000hm}), composite colored vector mesons  in technicolor models with colored technifermions (technirhos \cite{Farhi:1980xs,Hill:2002ap, Lane:2002sm}), and low-scale string resonances~\cite{Antoniadis:1990ew}.

An electrically neutral color-singlet vector boson ($Z^\prime$) often originates from extending the electroweak $U(1)$ or $SU(2)$ gauge group; for examples of $Z^\prime$ models, see Refs.~\cite{Langacker:2008yv, Leike:1998wr, Hewett:1988xc} and references therein. The $Z^\prime$ can have flavor universal~\cite{Senjanovic:1975rk, Georgi:1989ic, Georgi:1989xz} or flavor non-universal couplings to fermions~\cite{Muller:1996dj, Malkawi:1996fs, Chivukula:1994mn}; the latter happens when the gauge group for the $\zp$ does not commute with the $SU(2)_L$ of the standard model. While a typical $\zp$ can couple to leptons as well as quarks, it is possible (see, e.g.,~\cite{Harris:1999ya}) to have a $\zp$ that does not decay to charged leptons  and must be probed via its hadronic channels such as a dijet final state. In this article, we are interested in $\zp$ bosons of this kind (denoted as ``leptophobic''), because one of them could appear as a dijet resonance without any corresponding dilepton signature.

A coloron ($C$) or a $\zp$ manifesting as a dijet resonance is produced at hadron colliders via quark-antiquark annihilation%
\footnote{The resonances corresponding to these particles are not produced by gluon fusion: the $\zp$ is not colored and the coloron does not couple to gluon pairs (except very weakly at one loop and higher orders \cite{Chivukula:2013xla}).}%
. The interaction of a $C$ with the SM quarks $q_i$ is described by
	\beq
		\mathcal{L}_C  = i g_{QCD} C_\mu^a \sum_{i=u,d,c,s,t,b}
		\bar{q}_i\gamma^\mu t^a \left( g_{C_L}^i P_L + g_{C_R}^i P_R \right) q_i , \\
	\label{eq:colcoupl}
	\eeq
where $t^a$ is an $ \text{SU}(3) $ generator, while $g_{C_L}^i$ and $g_{C_R}^i$ denote left and right chiral coupling strengths (relative to the strong coupling $g_{QCD}$) of the color-octet to the SM quarks. The projection operators have the form $P_{L,R} = (1 \mp \gamma_5)/2$ and the quark flavor index runs over $ i=u,d,c,s,t,b.$ Similarly, the interactions of a leptophobic $\zp$ with the SM quarks are given by
	\beq
		\mathcal{L}_{\zp}  = i g_w  \zp_\mu \sum_{i=u,d,c,s,t,b}
		\bar{q}_i \gamma^\mu\left( g_{\zp_L}^i P_L + g_{\zp_R}^i P_R \right) q_i,
	\label{eq:zpcoupl}
	\eeq
where $g_{\zp_L}^i$ and $g_{\zp_R}^i$ denote left and right chiral coupling strengths of the leptophobic $\zp$ to the SM quarks, relative to the weak coupling $g_w = e/\sin\theta_W$.

In this analysis, we limit ourselves to vector resonances with flavor-universal couplings to quarks; that is, the couplings of each vector resonance to every quark, regardless of generation or chirality, is the same.  This simplifies the calculation of the widths.  For the multi-TeV resonances that are not yet excluded by experiment, the total decay width for a heavy coloron is
	\beq
		\Gamma_C = \frac{\alpha_s}{2} M_C g_C^2,
	\label{eq:colwiduniv}
	\eeq
and for a leptophobic $Z^\prime$ is

	\beq
		\Gamma_\zp = 3 \alpha_w M_\zp g_\zp^2\,,
	\label{eq:zpwiduniv}
	\eeq
where $g_{C/\zp}^2 = \left( g_{{C/\zp}_L}^2 + g_{{C/\zp}_R}^2\right)$ denotes the resonance's flavor-universal coupling to quarks.

\section{The Color Discriminant Variable}
\label{sec:coldis}
In this section, we review the color discriminant variable, as first presented in~\cite{Atre:2013mja}. Then we introduce a more general formulation of these ideas, drawing on Ref. \cite{Harris:2011bh} and use it to evaluate $D_{col}$ for various diquark states.

A scalar or vector resonance coupled to quarks in the standard model can be abundantly produced at a hadron collider of sufficient energy. Then it decays to a final state of simple topology: a pair of jets, top quarks, or bottom quarks, both of which are highly energetic and clustered in the central region of the detector. In a large data sample, a resonance with a relatively small width will appear as a distinct bump over a large, but exponentially falling, QCD background. These features make the hadronic decay channels favorable for discovery.

Searches for new particles currently being conducted at the LHC are focused on resonances having a narrow width. So one can expect that if a new dijet resonance is discovered, the  dijet cross section, mass,  and width of the resonance will be measured. These three observables are exactly what is needed to construct the color discriminant variable \cite{Atre:2013mja}, as defined in (\ref{eq:dcol}) that can distinguish between resonances of differing color charges.

\subsection{Review of the original formulation}
\label{subsec:flav-univ}

Let us review the idea behind the color discriminant variable, using a flavor-universal color-octet ($C$)  and color-singlet ($\zp$) vector resonances as an illustration. Throughout this article we will work in the limit of sufficiently small width ($\Gamma/M \ll 1$) such that the dijet cross section for a process involving a vector resonance $V$ can be written, using the narrow-width approximation, as:
	\beq
		\sigma_{jj}^V \equiv \sigma(pp \rightarrow V \rightarrow jj) \simeq \sigma(pp \to V) Br(V \to jj),
	\label{eq:genxsec}
	\eeq
where $\sigma(pp \to V)$ is the cross section for producing the resonance. Note that $Br(V \to jj)$ is the boson's dijet branching fraction, which equals $4/6$ for a flavor-universal vector resonance that is heavy enough to decay to all six quark flavors.   As discussed in Ref. \cite{Atre:2013mja}, incorporating the respective expressions for the vector resonance widths (\ref{eq:colwiduniv}, \ref{eq:zpwiduniv}) leads to the dijet cross sections for a coloron
	\beq
		\sigma_{jj}^C = 	\frac{16\pi^2}{9} \frac{\Gamma_C}{M_C^3}  Br(C \to jj) \sum_{q=u,d,c,s} \tau \left[ \frac{d{\cal L}^{q\bar{q}}}{d\tau}\right] _{\tau = \frac{M_C^2}{s}} ,
	\label{eq:colxsec}
	\eeq
and for a leptophobic $\zp$,
	\beq
		\sigma_{jj}^\zp = \frac{2\pi^2}{9} \frac{\Gamma_{\zp}}{M_{\zp}^3} Br(\zp \to jj) \sum_{q=u,d,c,s} \tau \left[ \frac{d{\cal L}^{q\bar{q}}}{d\tau}\right] _{\tau = \frac{M_{Z'}^2}{s}} \,.
	\label{eq:zpxsec}
	\eeq
The parton luminosity function\footnote{In our previous work \cite{Atre:2013mja,Chivukula:2014npa,Chivukula:2014pma}, involving only collisions of  non-identical partons $q$ and $\bar{q}$, we used the notation $W_{q}$ to refer to the parton luminosity: $$W_{q}(M_V) = 2\pi^2 \frac{M_V^2}{s} \int_{M_V^2/s}^{1} \frac{dx}{x}
			\left[ f_q\left(x, \mu_F^2\right) f_{\bar{q}}\left( \frac{M_V^2}{sx}, \mu_F^2 \right) +
			f_{\bar{q}}\left(x, \mu_F^2\right) f_q\left( \frac{M_V^2}{sx}, \mu_F^2 \right) \right] = 2\pi^2 \tau \left[ \frac{d{\cal L}^{q\bar{q}}}{d\tau}\right] _{\tau = \frac{M_V^2}{s}}.$$}
 $\tau d{\cal L}/d\tau$ for production of the vector resonance with mass $M_V$ via collisions of partons $i$ and $k$ at the center-of-mass energy squared $s$, is defined by
	\beq
	\tau \left[ \frac{d{\cal L}^{ik}}{d\tau}\right] \equiv 
	\frac{\tau}{1 + \delta_{ik}} \int_{\tau}^{1} \frac{dx}{x}
			\left[ f_i\left(x, \mu_F^2\right) f_k\left( \frac{\tau}{x}, \mu_F^2 \right) +
			f_{k}\left(x, \mu_F^2\right) f_i\left( \frac{\tau}{x}, \mu_F^2 \right) \right]  \,,
	\label{eq:lumi-fun}
	\eeq
where $f_{i}\left(x,\mu_F^2\right)$ is the parton distribution function at the factorization scale $\mu_F^2$. Throughout this article, we set the factorization scale equal to the resonance mass -- for vector resonances,  $\mu_F^2 = M_{C,Z'}^2$.

The fact that the overall coupling strength can be represented as the ratio of observables $\Gamma_V/M_V$, as shown in Eqs.~(\ref{eq:colwiduniv}, \ref{eq:zpwiduniv}), motivates the definition of the color discriminant variables:
	\bea
		\dcol^C &=&  \frac{M_C^3}{\Gamma_C} \sigma_{jj}^C= \frac{16\pi^2}{9} \left[ Br(C \to jj)  \sum_{q=u,d,c,s} \tau \left[ \frac{d{\cal L}^{q\bar{q}}}{d\tau}\right] _{\tau = \frac{M_C^2}{s}}  \right]
	\label{eq:coldcol}\\
		\dcol^\zp &=& \frac{M_\zp^3}{\Gamma_\zp} \sigma_{jj}^\zp= \frac{2\pi^2}{9} \left[ Br(\zp \to jj)  \sum_{q=u,d,c,s} \tau \left[ \frac{d{\cal L}^{q\bar{q}}}{d\tau}\right] _{\tau = \frac{M_{Z'}^2}{s}}  \right]
	\label{eq:zpdcol}
	\eea
for the coloron and $\zp$, respectively.

The factors in the square brackets in Eqs. (\ref{eq:coldcol}, \ref{eq:zpdcol}) are the same for flavor-universal resonances having a particular mass; only the initial numerical factors differ. In other words, the difference between the values of color discriminant variables corresponding to the two types of flavor-universal vector resonances
	\beq
		\dcol^C = 8 \dcol^\zp
	\label{eq:factor8_dcol}
	\eeq
will help pinpoint the nature of the color structure of the discovered particle. Turning this argument around, a new resonance with a particular mass, dijet cross-section and value of $\dcol$ could correspond to a broader $\zp$ or narrower coloron.

Since $D_{col}$ depends only on observables ($\sigma_{jj}$, $\Gamma$, $M$) that will be measured as soon as a new narrow resonance is seen, it provides a way to immediately distinguish color-octet and color-singlet dijet resonances.

\subsection{Alternative Description}
\label{subsec:alt-des}

We now present an alternative formulation of the tree-level $s$-channel resonance cross section which makes the properties of the color-discriminant variable more transparent and makes $D_{col}$ easier to calculate for diverse types of resonances. Following Eq.~(44) of \cite{Harris:2011bh}, the spin- and color-averaged partonic tree-level $s$-channel cross section for the process $i + k \to R \to x + y$ is written
\begin{equation}
\hat{\sigma}_{ik\to R\to xy}(\hat{s}) = 16 \pi \cdot {\cal N} \cdot (1 + \delta_{ik}) \cdot
\frac{\Gamma(R\to ik) \cdot \Gamma(R\to xy)}
{(\hat{s}-m^2_R)^2 + m^2_R \Gamma^2_R} ~,
\end{equation}
where $(1 + \delta_{ik})$ accounts for the possibility of identical incoming partons\footnote{More specifically, for identical incoming partons $ik$ the partial width $\Gamma(R \to ik)$ has an extra factor of 1/2 in the integration over final state phase-space -- a factor which is not present for distinguishable partons and must be removed when relating the partial width to the partonic production cross-section. While this factor is not explicitly included in \cite{Harris:2011bh}, it matters only for identical incoming partons.}.  The factor ${\cal N}$ is a ratio of spin and color counting factors
\begin{equation}
{\cal N} = \frac{N_{S_R}}{N_{S_i} N_{S_k}} \cdot
\frac{C_R}{C_i C_k}~,
\end{equation}
where $N_S$ and $C$ count the number of spin- and color-states for initial state partons $i$ and $k$. In the narrow-width approximation, we also have
\begin{equation}
\frac{1}
{(\hat{s}-m^2_R)^2 + m^2_R \Gamma^2_R}
\approx \frac{\pi}{m_R \Gamma_R} \delta(\hat{s} - m^2_R)~.
\end{equation}

Integrating over parton densities, and summing over incoming partons, as well as the outgoing partons that produce hadronic jets ($jj$), we then find the tree-level hadronic cross section to be
\begin{equation}
\sigma_R =
16\pi^2 \cdot {\cal N} \cdot \frac{ \Gamma_R}{m_R} \cdot
\left(\sum_{xy = jj} BR(R\to xy)\right)
\left( \sum_{ik} (1 + \delta_{ik}) BR(R\to ik) \left[\frac{1}{s} \frac{d L^{ik}}{d\tau}\right]_{\tau = \frac{m^2_R}{s}}\right)~.
\end{equation}
Hence, for the color discriminant variable
\begin{equation}
D_{col} = \frac{\sigma_R \cdot m^3_R}{\Gamma_R}~,
\end{equation}
we find the general expression
\begin{equation}
D_{col}=  16\pi^2 \cdot {\cal N}  \cdot
\left(\sum_{xy=jj} BR(R\to xy)\right)
\left( \sum_{ik} (1 + \delta_{ik}) BR(R\to ik) \left[\tau \frac{d L^{ik}}{d\tau}\right]_{\tau = \frac{m^2_R}{s}}\right)~.
\label{eq:dcol-expression}
\end{equation}
This expression illustrates the dependence of the color discriminant variable on the properties of the incoming and outgoing partons, and can easily be applied to any narrow resonance.\footnote{Note that here we refer to the total resonance production cross section, while in practice there can be (small) corrections due to differing experimental acceptances for resonances of different spins \cite{Harris:2011bh}.}

Applying this approach to the flavor-universal coloron resonance considered above, we note: the coloron has $C_R = 8$ and $N_{S_R} = 3$; the incoming $xy$ and outgoing $ik$ states are a light quark $q = u,d,c,s$ and its anti-quark $\bar{q}$; each incoming quark has $N_{S_i} = N_{S_k} = 2$ and $C_i = C_k = 3$; the sum over outgoing branching ratios ($Br(C \to xy)$) is 4/6; and each incoming branching ratio is ($Br(C\to ik) = 1/6$).  Then Eq.~(\ref{eq:dcol-expression}) becomes
\beq
D_{col}^C=  \frac{32\pi^2}{27}
\left( \sum_{q=u,c,d,s} \left[\tau \frac{d L^{q\bar{q}}}{d\tau}\right]_{\tau = \frac{m^2_C}{s}}\right)~.
\label{eq:dcol-expression-C-alt}
\eeq
which is identical to the result in Eq.~(\ref{eq:coldcol}).

\subsection{$D_{col}$ for scalar diquarks}
\label{subsec:alt-des-diquark}

\subsubsection{Color-triplet diquarks}

The color-triplet weak-singlet scalar diquark $\omega_3$ is produced via collisions of $u$ and $d$ quarks and decays back to the same quark pair:  $pp \to u_L d_L \to \omega_3 \to u_L d_L$.  For this process, we may determine the value of the color discriminant variable directly from Eq.~(\ref{eq:dcol-expression}).  Noting that the diquark has $N_{S_R} =1 $ and $C_R = 3$; the incoming and outgoing quarks have $N_{S_i}=N_{S_k} = 2$ and $C_i = C_k = 3$; and there is only one incoming and one outgoing mode (so the summations are superfluous), we find:
\beq
D_{col}^{\omega_3} =  \frac{4\pi^2}{3}
\left[ \tau \frac{d {\cal L}^{ud}}{d\tau} \right]_{\tau = \frac{m^2_{\omega_3}}{s}}~.
\label{eq:dcol-expression-D-alt}
\eeq
Constructing $D^{\omega_3}_{col}$ from the cross-section (\ref{eq:D-triplet-xsec}), width (\ref{eq:D-triplet-width}) and mass of the diquark, using Eq.~(\ref{eq:dcol}), yields the same result.

For the other color-triplet weak-singlet diquark $\tilde{\omega}_3$, we find
\beq
D_{col}^{{\tilde{D\omega}}_3} = D_{col}^{\omega_3}\,.
\eeq
This follows either by noting that $\omega_3$ and $\tilde{\omega}_3$ have the same the cross-sections (\ref{eq:D-triplet-xsec}) and widths (\ref{eq:D-triplet-width}), or by realizing that the factors contributing to ${\cal N}$ in Eq.~(\ref{eq:dcol-expression}) are the same as for $\omega_3$.

\subsubsection{Color-sextet diquarks}

The color-sextet weak-singlet scalar diquark $\Phi_6$ is produced via collisions of two $u$ quarks and decays back to the same quark pair:  $pp \to u_R u_R \to \Phi_6 \to u_R u_R$.  We may determine the value of $\dcol$ in a straightforward way from Eq.~(\ref{eq:dcol-expression}).  Noting that the diquark has $N_{S_R} = 1$ and $C_R = 6$; the incoming and outgoing quarks have $N_{S_i}=N_{S_k} = 2$ and $C_i = C_k = 3$; and there is only one incoming or outgoing mode, we conclude that ${\cal N} = 1/6$.  Furthermore, the two initial state partons are identical, so $(1 + \delta_{ik}) = 2$.  We therefore find
\beq
D_{col}^{\Phi_6} =  \frac{16\pi^2}{3}
\left[ \tau \frac{d {\cal L}^{uu}}{d\tau} \right]_{\tau = \frac{m^2_{\Phi_6}}{s}}~.
\label{eq:dcol-expression-Phi6}
\eeq

The calculation for $\phi_6$ is essentially unchanged, since $\phi_6$ couples only to $d_R d_R$.  The factor ${\cal N}$ is the same as for $\Phi_6$ and there is just one incoming and one outgoing mode to consider.  The only factor that distinguishes the results is the luminosity function.
\beq
D_{col}^{\phi_6} =  \frac{16\pi^2}{3}
\left[ \tau \frac{d {\cal L}^{dd}}{d\tau} \right]_{\tau = \frac{m^2_{\phi_6}}{s}}~.
\label{eq:dcol-expression-phi6}
\eeq

The case of $\Delta_6$ is very similar: ${\cal N}$ remains the same and there is still only one incoming and one outgoing mode.  This time, however, the incoming mode involves non-identical partons $u_R d_R$, so that $(1 + \delta_{ik}) = 1$.  As a result,
\beq
D_{col}^{\Delta_{6}} =  \frac{8\pi^2}{3}
\left[ \tau \frac{d {\cal L}^{ud}}{d\tau} \right]_{\tau = \frac{m^2_{\Delta_{6}}}{s}}~.
\label{eq:dcol-expression-Delta6}
\eeq

Finding $\dcol$ for the $\delta_6$ state takes a little more thought since there are two decay modes to consider: $\delta_6 \to us$ and $\delta_6 \to cd$, each with branching ratio $1/2$.  The factor ${\cal N}$ is the same as for the other sextet states.  The sum of the outgoing ($xy$) dijet branching ratios is still 1 since the $\delta_6$ does not decay to heavy flavors. As with $\Delta_6$, the incoming partons are distinct from one another. But there are now two luminosity functions to keep track of in $\dcol$:
\beq
D_{col}^{\delta_6} =  \frac{4\pi^2}{3}
\left[ \tau \frac{d {\cal L}^{us}}{d\tau}  +   \tau \frac{d {\cal L}^{cd}}{d\tau}  \right]_{\tau = \frac{m^2_{\delta_6}}{s}}~.
\label{eq:dcol-expression-delta6}
\eeq

Finally, constructing $D_{col}$ for each sextet state from its cross-section, width and mass (as given in Section~\ref{sec:sextets}) and using Eq.~(\ref{eq:dcol}) yields the same result as we have just derived.

\section{Diquark Accessibility at the 14 TeV LHC}
\label{sec:paramspace}

After a narrow dijet resonance has been discovered, one uses the measurements of three observables; dijet cross section, mass, and total decay width to evaluate $\dcol$ via Eq.~(\ref{eq:dcol}).  At the same time, one can use Eq.~(\ref{eq:dcol-expression}) to compare $\dcol$ with the predictions for various classes of dijet resonances.

In this section, we describe the region of diquark parameter space to which the method is applicable.\footnote{A detailed discussion for colorons and $Z'$ bosons was presented in~\cite{Atre:2013mja,Chivukula:2014npa}.} In this region the resonance has not already been excluded by the current searches, can be discovered at the $5\sigma$ level at the LHC 14 TeV after statistical and systematic uncertainties are taken into account, and has a total width that is measurable and consistent with the designation ``narrow''.

\begin{figure}[h]
{
\includegraphics[width=0.49\textwidth, clip=true]{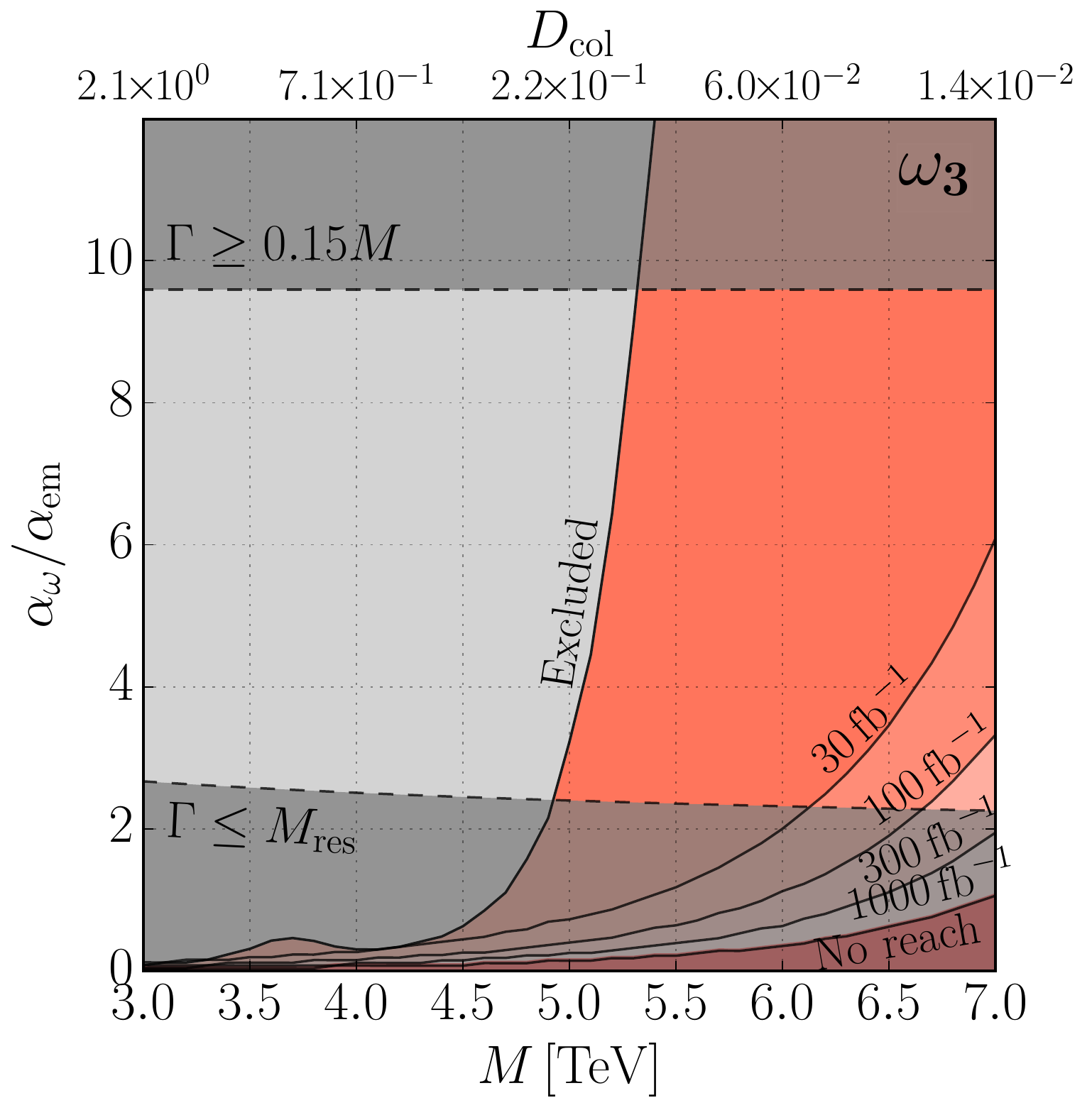}
}
\caption{\small 
Viable parameter (here $\alpha_\omega = \lambda^2/4\pi$, and is plotted relative to the electromagnetic coupling $\alpha_{em}$) space, the salmon-shaded area to the right of the curve labeled ``excluded", for the color triplet diquark $\omega_3$, which couples to $u_L d_L$; the viable parameter space for diquark $\tilde{\omega}_3$ is equivalent.  The curve at lower right labeled ``$30\, \ifb$'' delimits the region accessible to a $5\sigma$ discovery at the 14 TeV LHC with that integrated luminosity; the curves below it show how higher integrated luminosities (respectively, from above, 100, 300, 1000 $\ifb$) would increase the reach.  The region in which $D_{col}$ can be measured lies below dashed curve line where the resonance width equals 15\% of its mass and above the dashed curve where the resonance width equals the mass resolution of the detector is amenable; areas where the resonance is too broad or too narrow have been given a cloudy overlay.  The lowest red-shaded region lies beyond the reach of 1000 $\ifb$ of data.}
\label{fig:param_space_triplet}
\end{figure}

\begin{figure}[h]
{
\includegraphics[width=0.49\textwidth, clip=true]{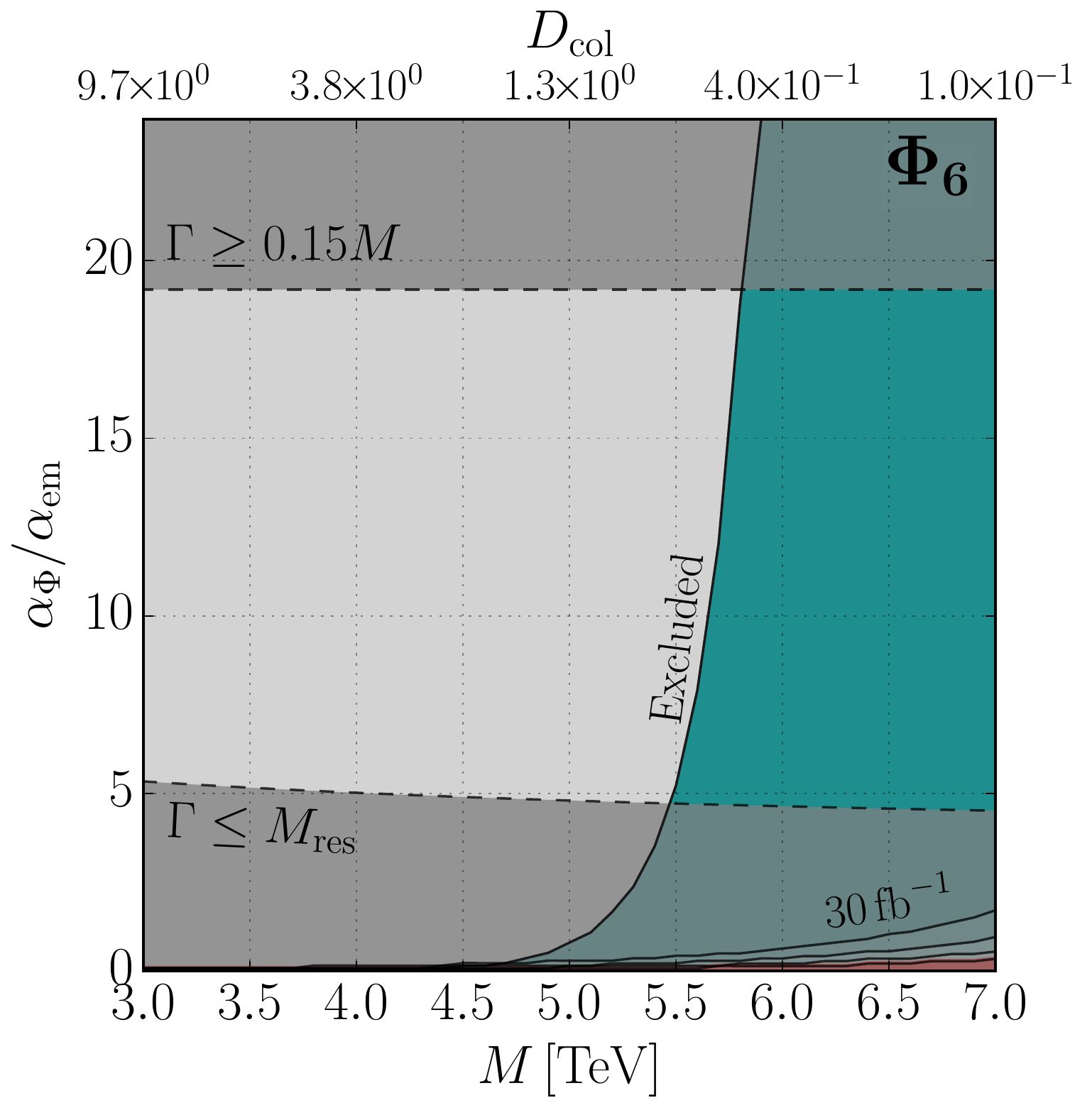}
\includegraphics[width=0.49\textwidth, clip=true]{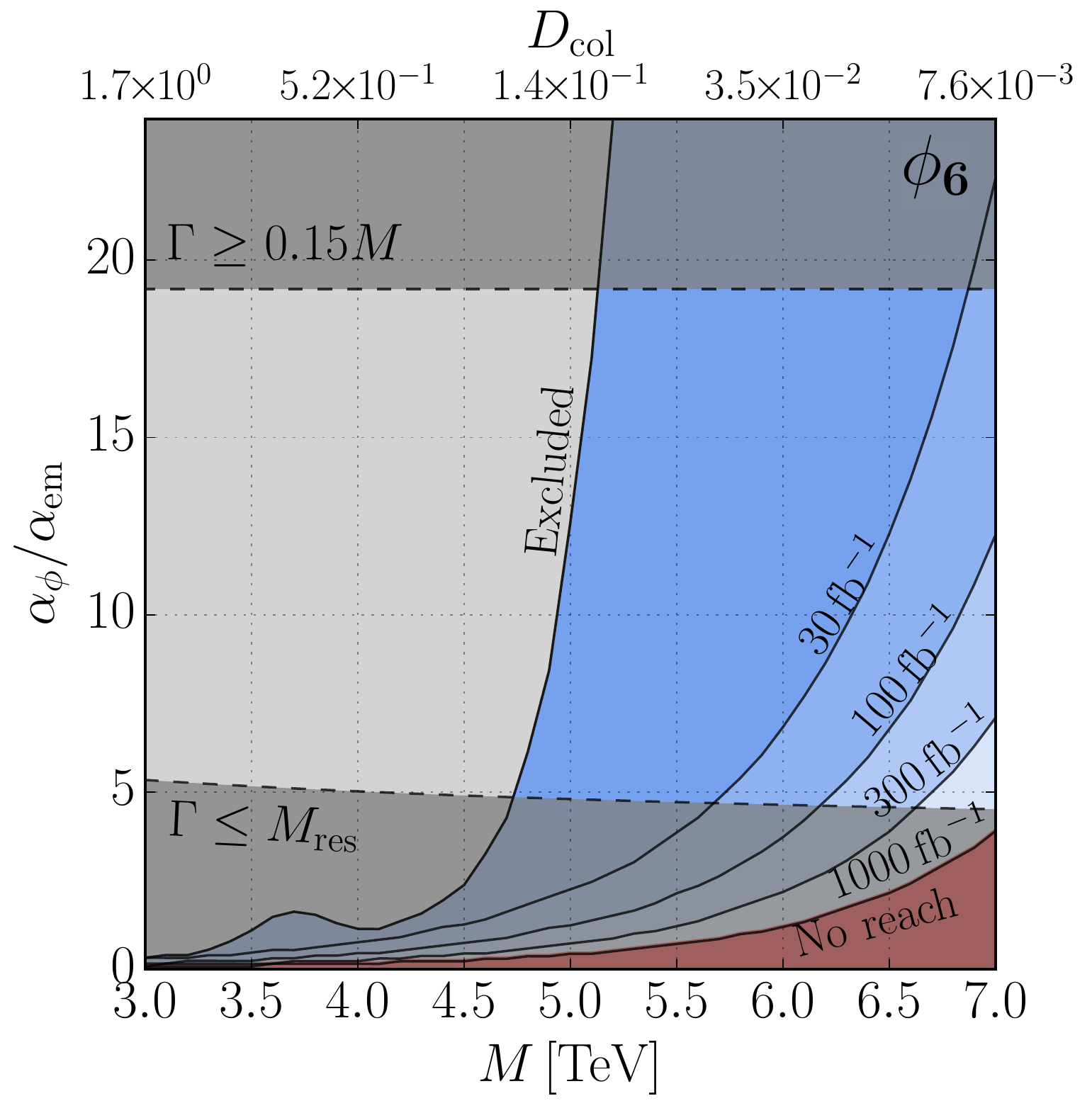}
}
\caption{\small 
Viable parameter space for color sextet diquarks: (Left) teal-shaded area to the right of the ``excluded" curve is for $\Phi_6$, which couples to $u_R u_R$,  (Right) blue-shaded area to the right of the  ``excluded" curve is for $\phi_6$, which couples to $d_R d_R$.  Other features of the figure are as in Fig.~\ref{fig:param_space_triplet}. The red area at far right labeled ``no reach" is not accessible even with $1000\, \ifb$ of integrated luminosity. Here $\alpha_{\Phi,\phi} = \lambda^2_{\Phi,\phi}/4\pi$.
}
\label{fig:param_space_sextet_phi}
\end{figure}

\begin{figure}[h]
{
\includegraphics[width=0.49\textwidth, clip=true]{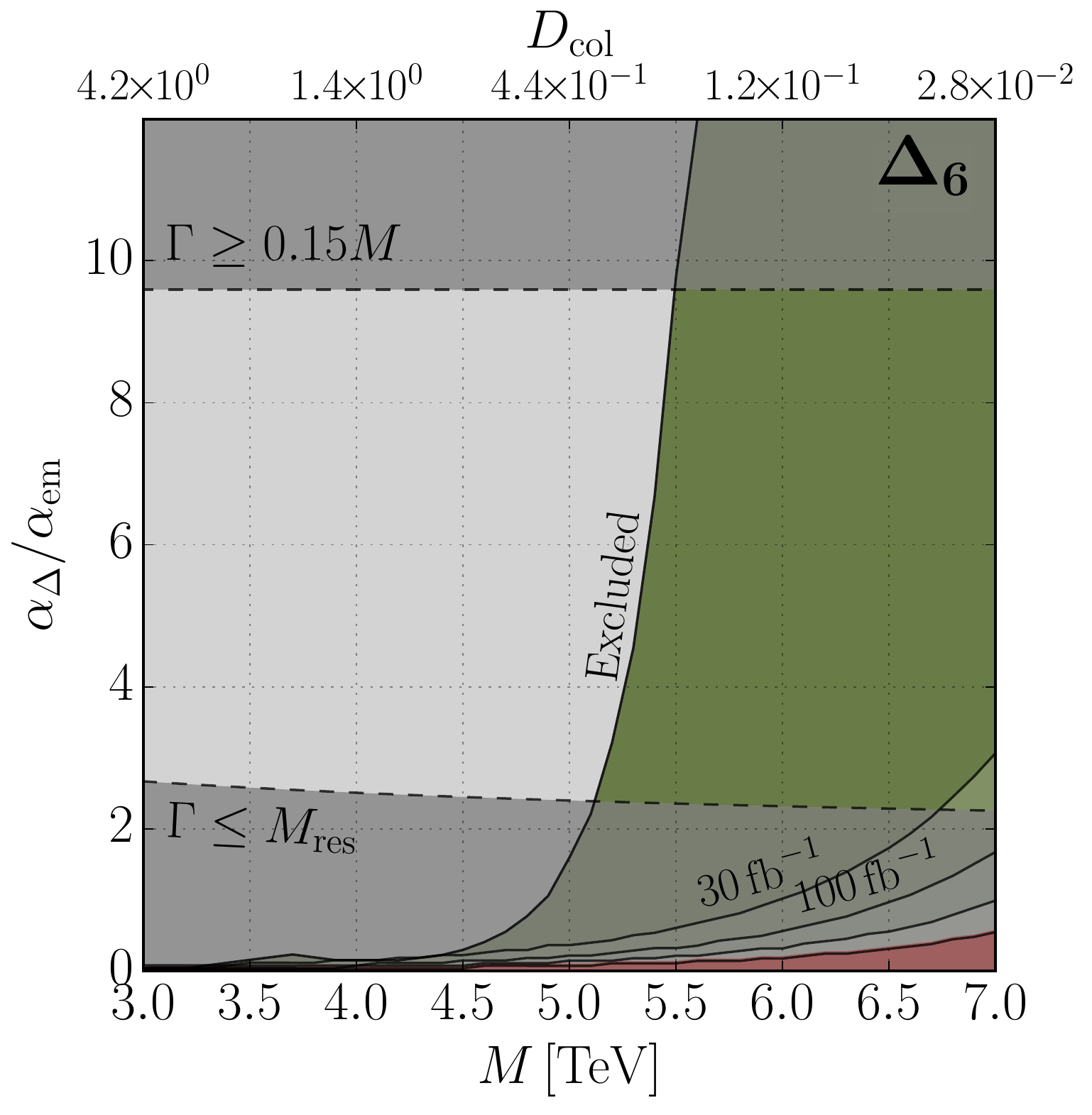}\includegraphics[width=0.49\textwidth, clip=true]{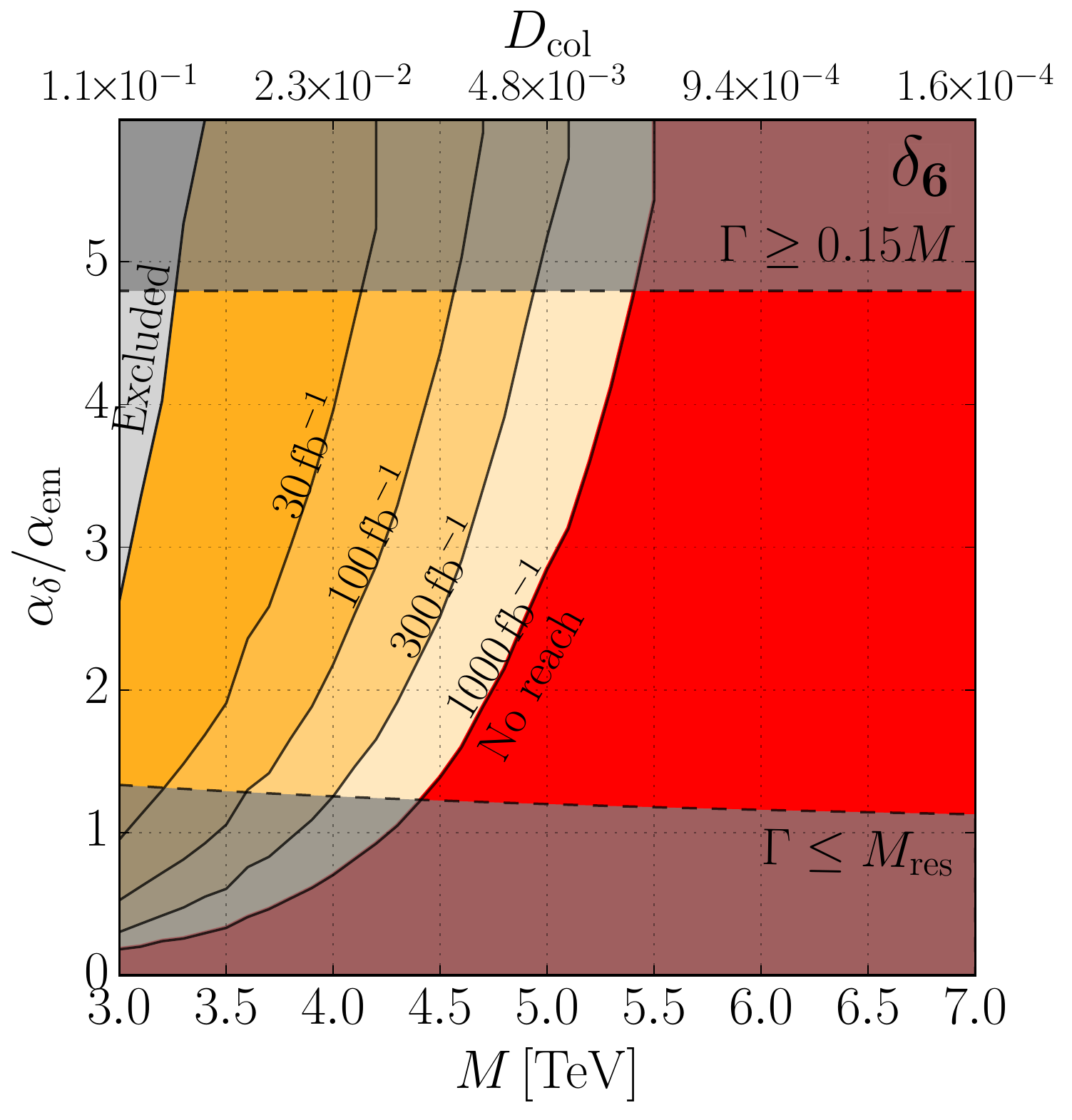}
}
\caption{\small 
Viable parameter space for color sextets diquarks: (Left) green-shaded area to the right of the ``excluded" curve is for $\Delta_6$, which couples to $u_L d_L$, (Right) yellow-shaded area to the right of the ``excluded" curve is for $\delta_6$, which couples to $(u_R s_R - c_R d_R)$.  Other features of the figure are as in Fig.~\ref{fig:param_space_triplet}.  The red area at far right labeled ``no reach" is not accessible even with $1000\, \ifb$ of integrated luminosity.
Here $\alpha_{\Delta,\delta} = \lambda^2_{\Delta,\delta}/4\pi$.
}
\label{fig:param_space_sextet_delta}
\end{figure}

One may deduce the current exclusion limits on diquark resonances using the limits on the production cross section times branching ratio ($\sigma \times Br(jj)$) from the (null) searches for narrow-width resonances carried out by the ATLAS and CMS collaborations~\cite{ATLAS:2012pu, Chatrchyan:2013qha, CMS:kxa} at $\sqrt{s} = 8\,\tev$. We use the most stringent constraint, which comes from CMS~\cite{CMS:kxa}. The exclusion limit for a given state is provided in the form of $\sigma\times Br(jj) \times (\mathrm{Acceptance})$.  For the $C$ and $\zp$ we have estimated~\cite{Atre:2013mja}  the acceptance of the detector for each value of the resonance mass by comparing, within the same theoretical model, the value of $\sigma \times Br(jj)$ that we have calculated with the value of $\sigma \times Br(jj) \times (\mathrm{Acceptance})$ provided by CMS. The acceptance is a characteristic of properties of the detector and kinematics, the latter being the same for a coloron and $\zp$ to leading order; thus we used throughout our vector boson analysis the acceptance deduced from such a comparison made within a sequential $\zp$ model. For scalar diquarks, we use a constant acceptance $0.6$ for a resonance with isotropic decays, as indicated by the authors of \cite{Khachatryan:2015sja}. The excluded region of parameter space is displayed in pale gray to the left of the curve labeled ``excluded'' in each plot in in Figs.~\ref{fig:param_space_triplet}, \ref{fig:param_space_sextet_phi}, and  \ref{fig:param_space_sextet_delta}.

Sensitivity to a dijet resonance in future LHC experiments with $\sqrt{s}=14\,\tev$ depends on the knowledge of QCD backgrounds, the measurements of dijet mass distributions, and statistical and systematic uncertainties. CMS
~\cite{Gumus:2006mxa} has estimated the limits on $\sigma \times Br(jj) \times (\mathrm{Acceptance})$ that will be required in order to attain a $5\sigma$ discovery at CMS with integrated luminosities up to $10\,\ifb$, including both statistical and systematic uncertainties. We obtain the acceptance for CMS at $\sqrt{s} = 14\,\tev$ in the same manner as described in the previous paragraph. The sensitivity for dijet resonance discovery with $10\,\ifb$ of data is then scaled to the integrated luminosities $\mathcal{L} = 30,\,100,\,300,\,1000\,\ifb$ considered in our studies (assuming that the systematic uncertainty scales with the squared root of integrated luminosity).  Since the original limits were estimated for resonances with masses up to $5\,\tev$, we have extrapolated the sensitivity to $7\,\tev$. The predicted diquark discovery reaches for these luminosities are shown in shades of dark pink for color-triplet diquarks $\omega_3$
and shades of  teal ($\Phi_6$), light blue ($\phi_6$), dark green ($\Delta_6$), and yellow ($\delta_6$) for color-sextet diquarks, in Figs. \ref{fig:param_space_triplet} \ref{fig:param_space_sextet_phi}, and  \ref{fig:param_space_sextet_delta}.

The total decay width also constrains the absolute values of the coupling constants. On the one hand, experimental searches are designed for narrow-width dijet resonances; hence their exclusion limits are not applicable when the resonance is too broad, which translates to about $\Gamma/M = 0.15$ as the upper limit~\cite{Bai:2011ed, Haisch:2011up, Harris:2011bh}. On the other hand, the appearance of (intrinsic) total decay width in the expression for the color discriminant variable requires that the width be accurately measurable; width values smaller than the experimental dijet mass resolution, $M_\mathrm{res}$, cannot be distinguished. The region of parameter space that meets both constraints and is relevant to our analysis is shown in Figs.~\ref{fig:param_space_triplet}, \ref{fig:param_space_sextet_phi}, and  \ref{fig:param_space_sextet_delta} as the region between the two dashed horizontal curves labeled $\Gamma \ge 0.15 M$ and $\Gamma \le M_\mathrm{res}$. Regions where the width is too broad or too narrow are shown with a cloudy overlay to indicate that they are not accessible via our analysis.

Statistical and systematic uncertainties on dijet cross section, mass, and intrinsic width of the resonance will play a key role in determining how well $\dcol$ can discriminate between models at the LHC with $\sqrt{s} = 14\,\tev$.  While the actual values of the systematic uncertainties at the LHC with $\sqrt{s} = 14\,\tev$ will be obtained only after the experiment has begun, we have previously discussed estimates of the uncertainties in \cite{Atre:2013mja,Chivukula:2014npa,Chivukula:2014pma}.  In particular, we reviewed estimates of the effect of systematic uncertainties in the jet energy scale, jet energy resolution, radiation and low mass resonance tail and luminosity on the dijet cross section at the 14 TeV LHC from  Ref.~\cite{Gumus:2006mxa} and discussed how this, combined with the dijet mass resolution would impact measurements of $\dcol$. We also estimated the uncertainty on $\dcol$ due to PDF uncertainties by using the CT10NLO PDF set from the CTEQ collaboration \cite{Lai:2010vv}; see Appendix~\ref{sec:PDFappendix} for details.  Overall, it appears that uncertainties of 20 - 50\% should be achievable and with that level of accuracy the color discriminant variable should be a useful tool for distinguishing among dijet resonances.


\section{Distinguishing Diquarks from Vector Bosons}
\label{sec:result-dcols}
We are now ready to illustrate how the color discriminant variable $\dcol$  may be used to distinguish whether a newly discovered dijet resonance is a scalar diquark, as opposed to a coloron or a leptophobic $\zp$.  As previously mentioned, we will focus on resonances having masses of $3-7\,\tev$ at the $\sqrt{s} = 14\,\tev$ LHC with integrated luminosities up to $1000\,\ifb$. The values of $\dcol$ and other observables have been evaluated using the uncertainties discussed in \cite{Atre:2013mja,Chivukula:2014npa,Chivukula:2014pma} and the region of parameter space to which this analysis is applicable was identified in Section \ref{sec:paramspace}.

\begin{figure}[h]
{
\includegraphics[width=0.49\textwidth, clip=true]{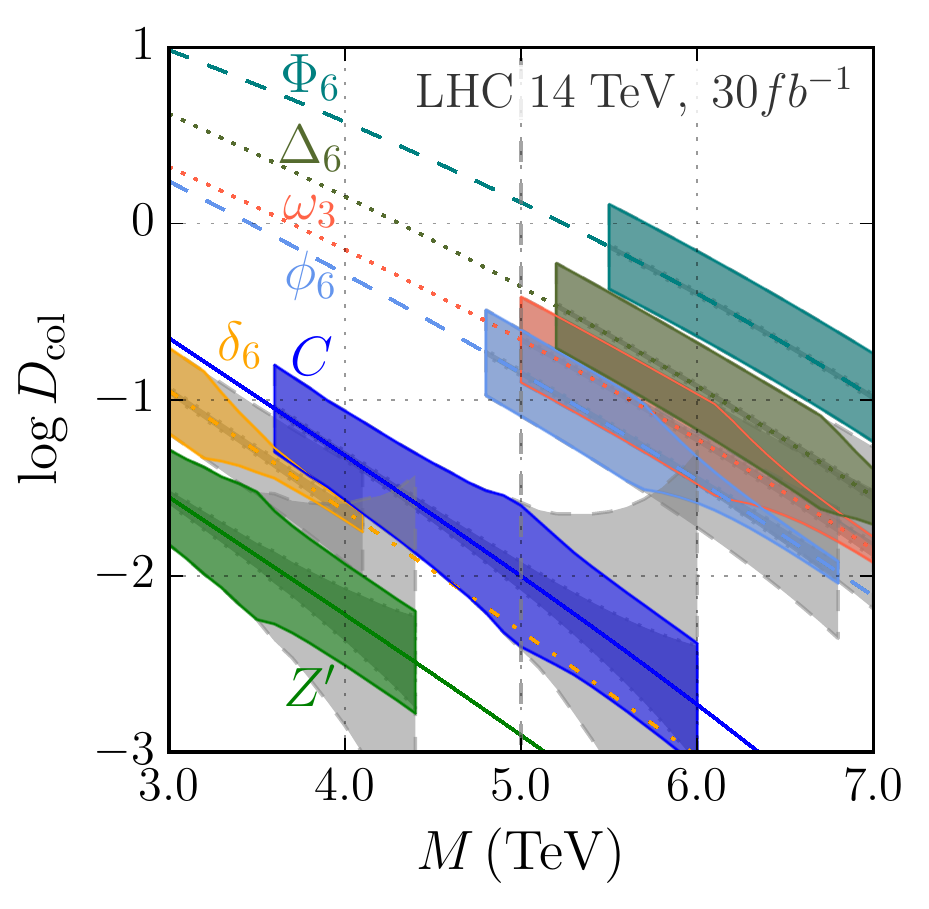}
\includegraphics[width=0.49\textwidth, clip=true]{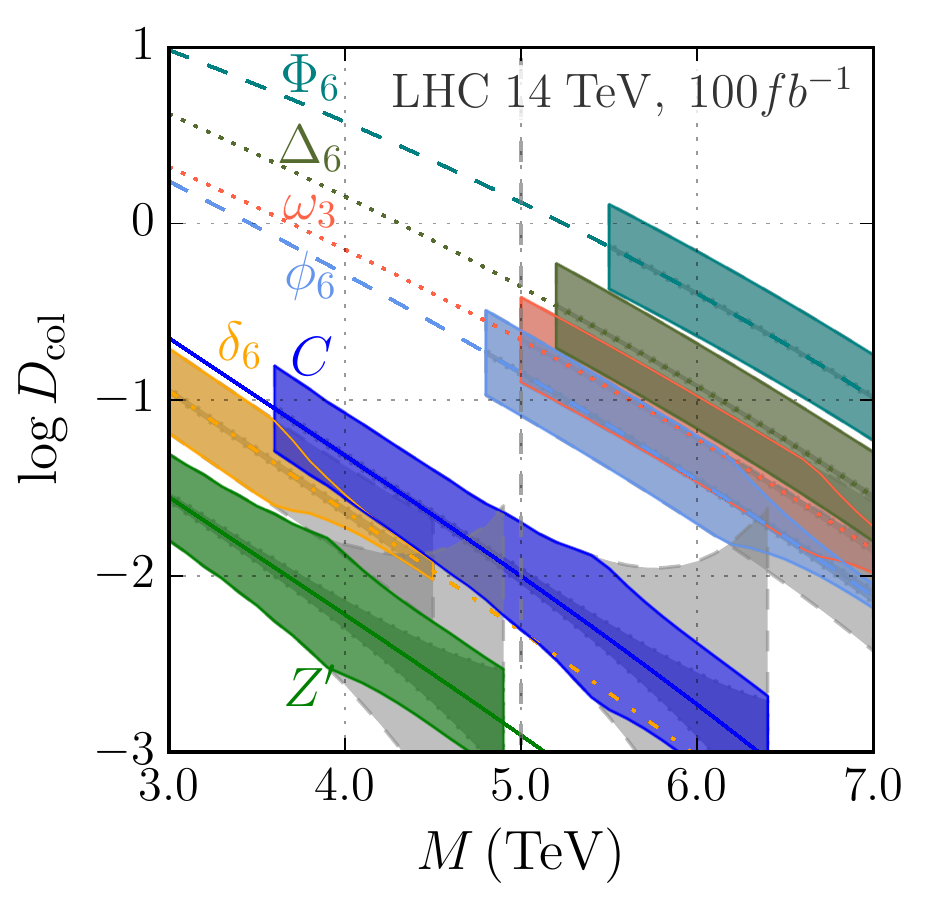}
}
\caption{\small
Color discriminant variables calculated, as described in Sec.~\ref{sec:paramspace}, using estimated systematic and statistical uncertainties for mass, total width, and dijet cross section for the integrated luminosities $30\,~\ifb$ (Left) and $100\,~\ifb$ (Right).  The central value of $\dcol$ for each particle is shown as, from top to bottom: $\Phi_6$ (dashed red), $\Delta_6$ (dotted black), $\omega_3$ (dotted red), $\phi_6$ (dashed black), $C$ (solid blue), $\delta_6$ (dotted-dash yellow), $\zp$ (solid green). The uncertainty in the measurement of $\dcol$ due to the uncertainties in the measurement of the cross section, mass and width of the resonance is indicated by gray bands. The outer (darker gray) band corresponds to the uncertainty in $\dcol$ when the width is equal to the experimental mass resolution i.e. $\Gamma = M_\text{res}$. The inner (lighter gray) band corresponds to the case where the width $\Gamma= 0.15M$. Resonances with width $M_\text{res} \leq \Gamma \leq 0.15M$ will have bands whose widths fall between the outer and inner gray bands. The horizontal extent of the colored band for each state indicates the region in parameter space where the particle has not been excluded by current searches and has the potential to be discovered at the $5\sigma$ level at the LHC 14 TeV after statistical and systematic uncertainties are taken into account. The limits and reaches for resonances with masses above $5\,\tev$ have been extrapolated, as mentioned in the text; as a reminder, the extrapolated region lies to the right of a vertical dashed line. 
}
\label{fig:pcombined-dcols-sys-stat-errors_30_100}
\end{figure}

\begin{figure}[h]
{
\includegraphics[width=0.49\textwidth, clip=true]{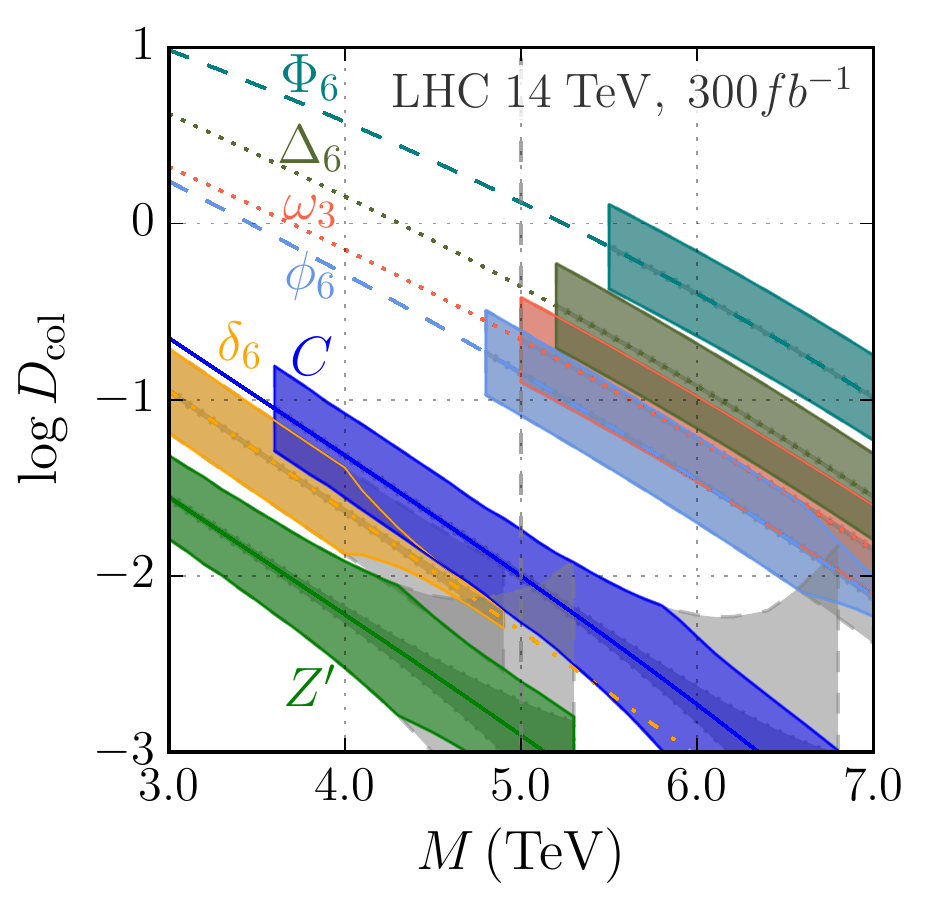}
\includegraphics[width=0.49\textwidth, clip=true]{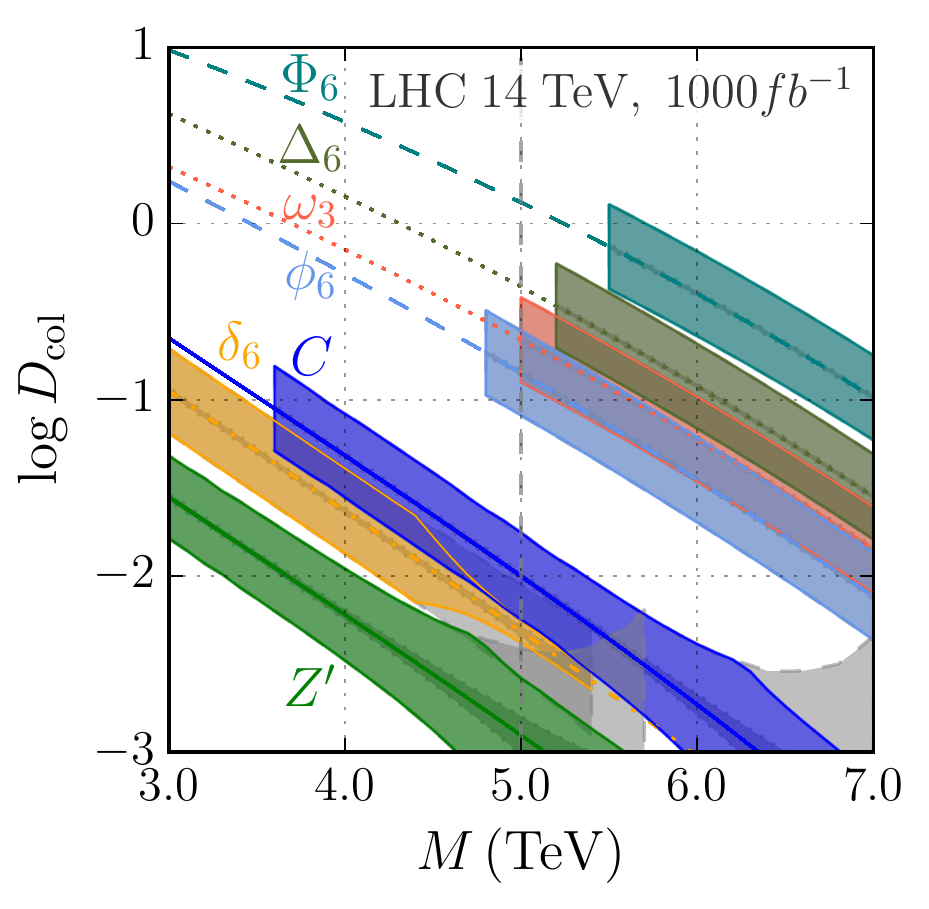}
}
\caption{\small 
Same as Fig.~\ref{fig:pcombined-dcols-sys-stat-errors_30_100}, but for integrated luminosities $300\,~\ifb$ (Left) and $1000\,~\ifb$ (Right).
}
\label{fig:pcombined-dcols-sys-stat-errors_300_1000}
\end{figure}

Figures~\ref{fig:pcombined-dcols-sys-stat-errors_30_100} and ~\ref{fig:pcombined-dcols-sys-stat-errors_300_1000} compare the value of $D_{col}$ as a function of resonance mass for several different resonances: colorons, $\zp$ bosons, the color-triplet diquark $\omega_3$ and the color-sextet diquarks $\Phi_6$, $\phi_6$, $\Delta_6$ and $\delta_6$.  Fig.~\ref{fig:pcombined-dcols-sys-stat-errors_30_100} focuses on integrated luminosities of 30 and 100 $\ifb$, while Fig.~\ref{fig:pcombined-dcols-sys-stat-errors_300_1000} displays results for 300 and 1000 $\ifb$.  In each plot, a given colored band shows the mass range in which the corresponding resonance is viable and accessible.  The appropriate exclusion limit in Figs.~\ref{fig:param_space_triplet}, \ref{fig:param_space_sextet_phi}, or \ref{fig:param_space_sextet_delta} delimits the left-hand edge of each band; the appropriate integrated luminosity curve from Figs.~\ref{fig:param_space_triplet}, \ref{fig:param_space_sextet_phi}, or  \ref{fig:param_space_sextet_delta} delimits the right-hand edge, beyond which there is not enough data to allow discovery at a given mass.  The width of each band relates to measurement uncertainties, as detailed in the figure caption.

For a given value of the dijet resonance mass, some types of resonance may already be excluded (e.g., a resonance found below about 3.4 TeV cannot be a coloron), while others may lie beyond the LHC's discovery reach at a given integrated luminosity, because too few events would be produced (e.g., a leptophobic $\zp$ will not be seen above about 4.8 TeV with only 100 $\ifb$).  But for any resonance mass between about 3.5 and 7 TeV, there are generally several dijet resonances that remain viable candidates For example, a resonance discovered at 4.0 TeV could be a $\zp$, coloron, or $\delta_6$, while one found at 5.2 TeV could be a $\Phi_6$, $\omega_3$ or coloron (or, with sufficient integrated luminosity, even a $\delta_6$ or a $\zp$).  

In many situations where a dijet resonance of a given mass discovered at LHC could correspond to more than one class of particle, measuring $D_{col}$ will suffice to distinguish among them.  A leptophobic $\zp$ would not be confused with any of the weak-singlet scalar diquarks (except, possibly, the $\delta_6$ near the top of the mass range for a given integrated luminosity).  Nor would any of the color-sextet diquarks be mistaken for one another.  The color-triplet diquark and the coloron are, likewise, distinct by this measure.

In other cases, the measurement of $\dcol$ will suffice to show that a resonance is a diquark, and yet may not have sufficient precision to determine which kind of diquark state has been found.  For instance, at masses of order 6 TeV, the sextet states are all distinct from one another, but the $\omega_3$ overlaps both the $\Phi_6$ and $\phi_6$. Measuring the color flow~\cite{Maltoni:2002mq,Kilian:2012pz,Curtin:2012rm,Gallicchio:2010sw} in the events may be of value here.  

In other cases, the measurement of $\dcol$ may leave us unsure as to whether a vector boson (coloron) or a color-sextet $\delta_6$ diquark has been discovered (e.g., at masses of order 3.5 TeV). In this case, measuring the angular distributions of the final state jets may assist in further distinguishing the possibilities \cite{Harris:2011bh}. 

\section{Discussion}
\label{sec:discussions}

The current run of the LHC has the potential to discover a new dijet resonance, opening the doors to an era of physics beyond the standard model.  The simple topology and large production rate for a dijet final state will not only promote discovery, but also aid in the determination of crucial properties of the new resonance. Because the color discriminant variable~\cite{Atre:2013mja}, $\dcol$, is constructed from measurements available directly after the discovery of the resonance via the dijet channel, namely, its mass, its total decay width, and its dijet cross section, this variable can be valuable in identifying the nature of a newly discovered state ~\cite{Atre:2013mja,Chivukula:2014npa,Chivukula:2014pma}.

In this work, we have extended the color discriminant variable technique in two directions. First, we have placed the theoretical discussion of the variable in more general language that shows its broader applicability and its relationship to the properties of the partons involved in production and decay of the resonance.  Second, we have shown that $\dcol$ may be used both to identify scalar diquark resonances as color triplet or color sextet states and to distinguish them from color-neutral or color-octet vector bosons.

In addition to the dijet resonances that we have discussed in detail in this work, other exist -- and they are also amenable to analysis via the color discriminant variable. One example would be states whose decay products include gluons: scalar resonances decaying to $gg$ and excited quarks decaying to $qg$.  As shown in~\cite{Chivukula:2014pma}, $\dcol$ can distinguish moderate width diquarks ($qq$) and vector bosons ($q\bar{q}$), whose decay products do not include gluons, from resonances whose decays do include final state gluons.  When the resonance is too narrow for $\dcol$ to be measurable, studying the jet energy profile of the final state jets can be a valuable alternative. Another example of additional dijet resonances would be diquarks that transform as weak triplets (cases XIII and XIV of \cite{Arnold:2009ay}), instead of being weak triplets like those studied here.  We have calculated the values of the color discriminant variable for all of these examples and presented them in Appendix~\ref{sec:Particleappendix} for comparison.  They fall within similar ranges to those discussed in detail above.

We hope the color discriminant variable will called upon soon to identify a new dijet resonance discovered at the LHC.

\begin{acknowledgments}
This material is based upon work supported by the National Science Foundation under Grant No. PHY-0854889. We wish to acknowledge the support of the Michigan State University High Performance Computing Center and the Institute for Cyber Enabled Research. PI is supported by Development and Promotion of Science and Technology Talents Project (DPST), Thailand.  EHS and RSC thank the Aspen Center for Physics and the NSF Grant \#1066293 for hospitality during the writing of this paper.
\end{acknowledgments}


\appendix

\section{Parton Luminosity Functions and Uncertainties}
\label{sec:PDFappendix}

This appendix illustrates the relevant parton luminosity functions, along with their uncertainties, for production of scalar diquarks and vector bosons.  We display these quantities, $\tau d{\cal L}^{ik}/d\tau$, as functions of resonance mass and show how they are related to the behavior of the central value of (and uncertainties in) $\dcol$ for a variety of dijet resonances.

In the left panel of Fig. \ref{fig:sum_pdf_dcol_ct10}, we show the parton luminosities corresponding to the initial states that produce the resonances studied in this paper. The right panel of Fig. \ref{fig:sum_pdf_dcol_ct10} shows $\dcol$ for the various resonances, as calculated using those parton luminosities. Uncertainties in the parton distribution functions, as calculated from the CT10NLO PDF set \cite{Lai:2010vv}, have been included; uncertainties from other sources are \textit{not} taken into account in this figure.  Note how the central values of the $\dcol$ curves correspond to those in Figs.~\ref{fig:pcombined-dcols-sys-stat-errors_30_100} and~\ref{fig:pcombined-dcols-sys-stat-errors_300_1000}.

\begin{figure}[h]
{
\includegraphics[width=0.49\textwidth, clip=true]{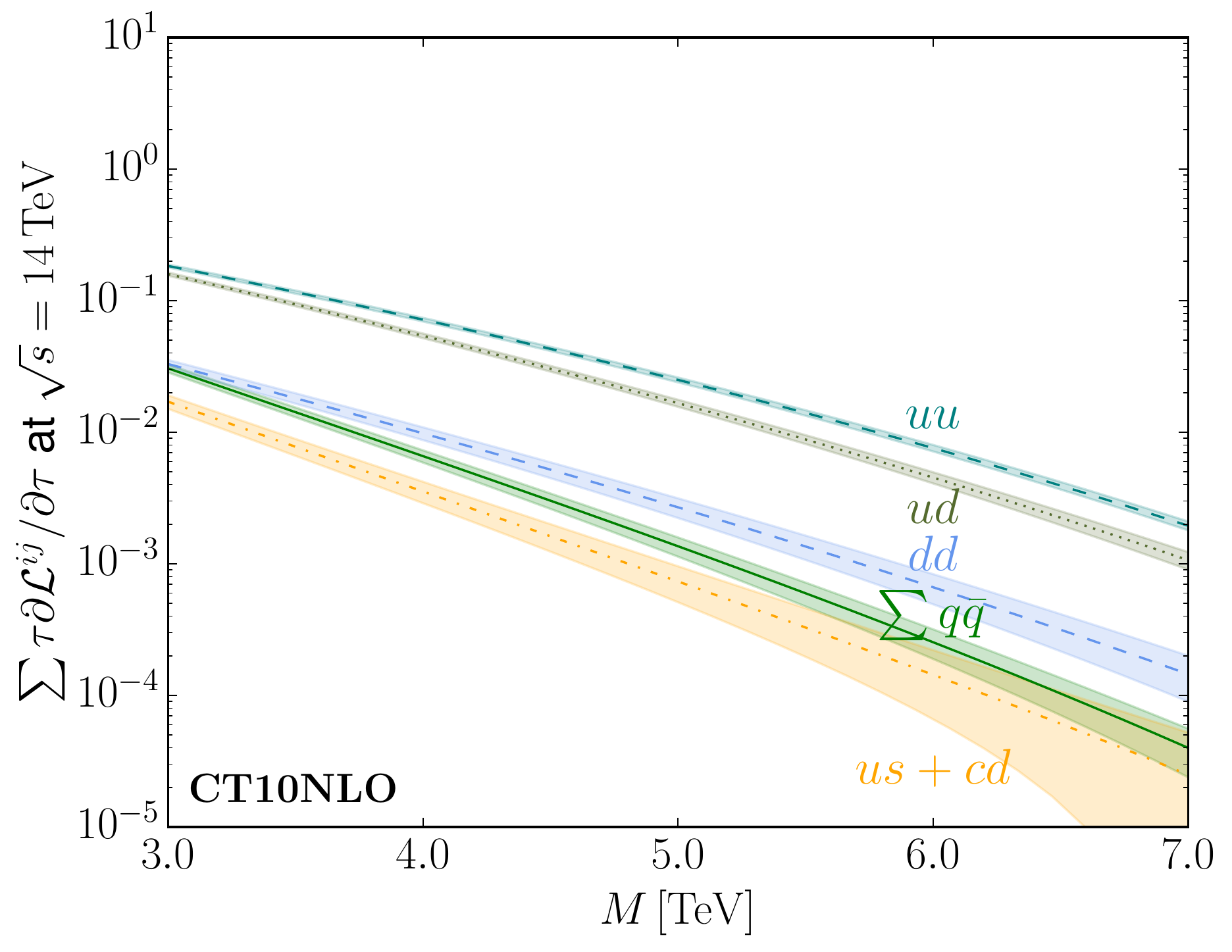}
\includegraphics[width=0.49\textwidth, clip=true]{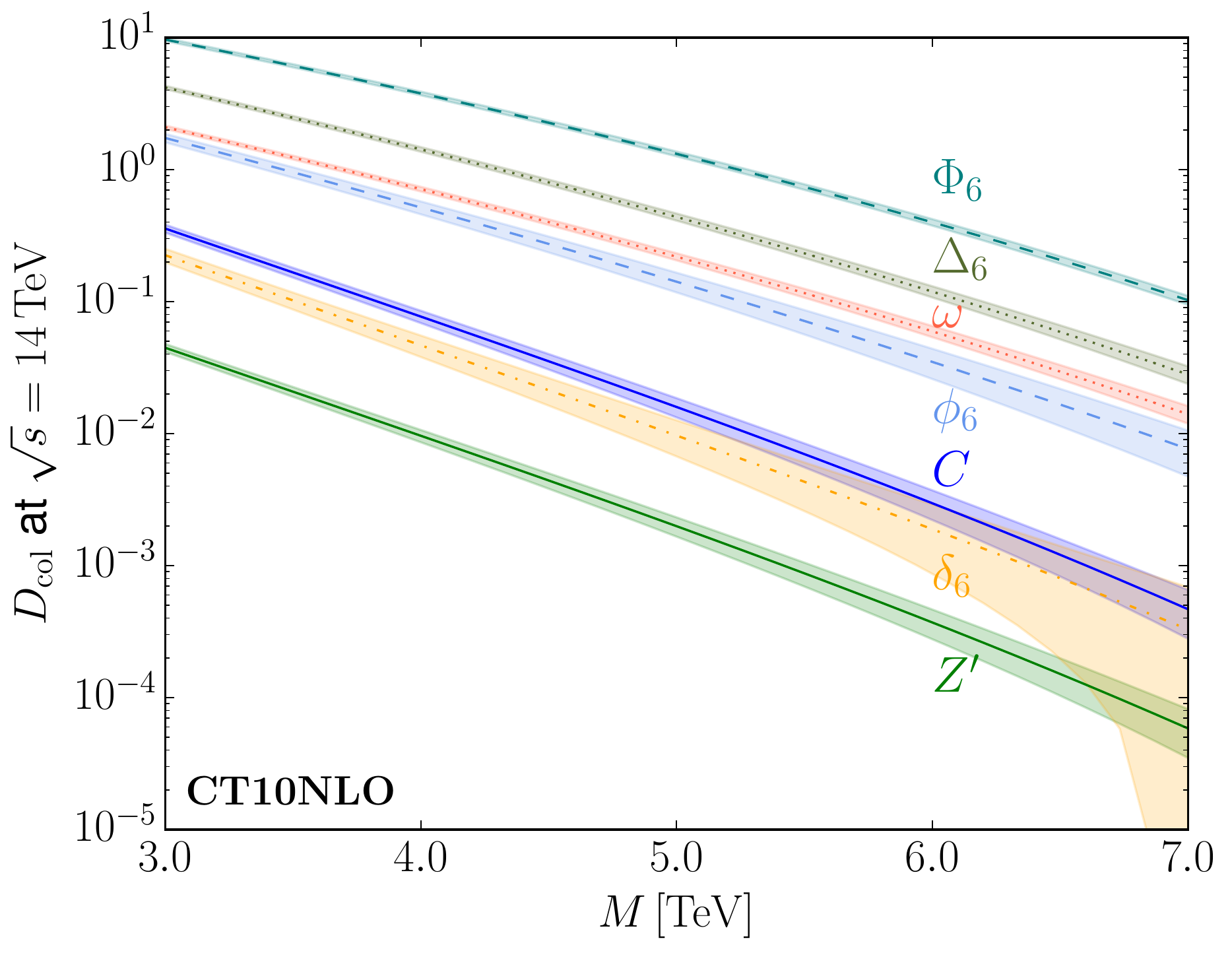}
}
\caption{\small
(a) Left: The sum of parton luminosity functions for different classes of initial states. From top to bottom: $uu$ (dashed red), $ud$ (dotted black), $dd$ (dashed black), $q\bar{q} = u\bar{u} + \ldots + b\bar{b}$ (solid green), $us+cd$ (dotted-dash yellow). (b) Right: Color discriminant variable curves for the particles discussed, including only the uncertainties due to the parton distribution functions. From top to bottom: $\Phi_6$ (dashed teal), $\Delta_6$ (dotted dark green), $\omega_3$ (dotted pink), $\phi_6$ (dashed blue), $C$ (solid blue), $\delta_6$ (dotted-dash yellow), $\zp$ (solid green). Each shaded band illustrates uncertainty due to the parton distribution functions, based on the CT10NLO PDF by the CTEQ Collaboration.
}
\label{fig:sum_pdf_dcol_ct10}
\end{figure}

\section{Color Discriminant Variable for Other Dijet Resonances}
\label{sec:Particleappendix}

\subsection{Weak-triplet scalar diquarks}

In addition to the weak-singlet scalar diquarks that have been the focus of this paper, there are also scalar diquarks transforming as weak triplets.   Both of these are charge $1/3$ states coupled to ${\bf\rm Q_L Q_L}$.  Each of them is anti-symmetric under exchange of Lorentz indices and symmetric under exchange of weak indices.

The first weak-triplet diquark is also a color triplet, making it anti-symmetric under $SU(3)_c$.  Hence, it must be anti-symmetric under flavor as well, making it a triplet under $SU(3)_{\rm Q_L}$.   The flavor combination coupling only to the two light generations is, therefore $\left[u_L c_L,\, (u_L s_L + c_L d_L)/\sqrt{2},\, d_L s_L\right]$.  Of these, it is the state coupled to $u_L s_L + c_L d_L$ that will be most readily produced at the LHC; this corresponds to case XIII in~\cite{Arnold:2009ay} and we will refer to it as $\chi_3$.  Based on all of this information, we find that the value of ${\cal N}$ is $1/12$.  The calculation of $\dcol$ proceeds similarly to that for the $\delta_6$ state (see discussion leading to Eq.~(\ref{eq:dcol-expression-delta6})), yielding:
\beq
D_{col}^{\chi_3} =  \frac{2\pi^2}{3}
\left[ \tau \frac{d {\cal L}^{us}}{d\tau}  +   \tau \frac{d {\cal L}^{cd}}{d\tau}  \right]_{\tau = \frac{m^2_{\chi_3}}{s}}~.
\label{eq:dcol-expression-chi3}
\eeq
In other words, because $\chi_3$ is a color-triplet but couples to the same mix of partons as $\delta_6$, its value of $\dcol$ is half that of $\delta_6$.

The second weak-triplet diquark is a color sextet, making it symmetric under exchange of color indices.  Hence it will also transform symmetrically under flavor -- specifically, it will be in a sextet under $SU(3)_{\bf\rm Q_L}$.  Of the six flavor combinations, the one coupled entirely to the first generation is $\left[ u_L u_L,\, (u_L d_L + d_L u_L)/\sqrt{2},\, d_L d_L\right]$.  The first component $\chi_6 \equiv u_L u_L$ will have the highest production rate at LHC; this corresponds to case XIV in~\cite{Arnold:2009ay}.  We find that ${\cal N} = 1/6$ and we note that the two incoming partons are identical.  Hence, the value of $\dcol$ is
\beq
D_{col}^{\chi_6} =  \frac{16\pi^2}{3}
\left[ \tau \frac{d {\cal L}^{uu}}{d\tau} \right]_{\tau = \frac{m^2_{\chi_6}}{s}}~.
\label{eq:dcol-expression-chi6}
\eeq
This is, not surprisingly, identical to the color discriminant variable for the $\Phi_6$ state that couples to $u_R u_R$.

\subsection{Dijet states coupling to gluons}

For completeness, we also present the formulae for the color discriminant variables for color-octet scalars (which couple to gluon pairs) and color-triplet excited quarks (which couple to $qg$).   In Ref.~\cite{Chivukula:2014pma}, we compared how $\dcol$ and the Jet Energy Profile could be used to distinguish these states from one another and from colorons at LHC.

We will designate a color-octet scalar coupling to gluon pairs by the symbol $\phi_8$.  For this state,  $C_{\phi_8} = C_g = 8$ and $N_{S_{\phi_8}} = 1$; also, $N_g$ = 2; hence, ${\cal N} = 1/32$.  Noting that the only decay of $\phi_8$ is to gluons, which are identical particles, we arrive at a color discriminant variable of
\beq
D_{col}^{\phi_8} =  {\pi^2}
\left[ \tau \frac{d {\cal L}^{gg}}{d\tau} \right]_{\tau = \frac{m^2_{\phi_8}}{s}}~.
\label{eq:dcol-expression-phi8}
\eeq
As illustrated in Fig. 2 of \cite{Chivukula:2014pma}, this falls between the value of $\dcol$ for a coloron and a $\zp$.

Calling an excited quark of a given flavor $q*$, and limiting ourselves to the quarks of the first generation, which are most strongly produced at the LHC, we note that the only decays are $u^* \to u g$ and $d^* \to d g$.  In either case, $N_{S_{q^*}} = N_{S_q} = N_{S_g} = 2$ and $C_{q^*} = C_q = 3$ while $C_g = 8$; as a result, ${\cal N} = 1/16$ for either excited quark.  Thus, we may write down the color discriminant variables as
\bea
D_{col}^{u^*} &=&  {\pi^2}
\left[ \tau \frac{d {\cal L}^{ug}}{d\tau} \right]_{\tau = \frac{m^2_{u^*}}{s}}\\
D_{col}^{d^*} &=&  {\pi^2}
\left[ \tau \frac{d {\cal L}^{dg}}{d\tau} \right]_{\tau = \frac{m^2_{d^*}}{s}}~.
\label{eq:dcol-expression-qstar}
\eea
As illustrated in Fig. 2 of \cite{Chivukula:2014pma}, this exceeds the value of $\dcol$ for a coloron.


\bibliography{diquark}
\bibliographystyle{apsrev}

\end{document}